\documentclass[10pt]{iopart}

\usepackage{amssymb}
\usepackage{graphicx}
\usepackage{epstopdf}
\usepackage{textcomp}
\usepackage{color}
\usepackage{filecontents}
\usepackage{cite}
\usepackage{url}
\usepackage{iopams} 
\expandafter\let\csname equation*\endcsname\relax
\expandafter\let\csname endequation*\endcsname\relax 
\usepackage{amsmath}
\usepackage{amsfonts} 

\begin{document}

\title[Quantum states of confined H$_2^+$]{Quantum states of confined hydrogen plasma species: Monte Carlo calculations}
\author{G. Micca Longo$^1$ , S. Longo$^{1,2,3}$ and D. Giordano$^4$}

\address{$^1$ CNR-Nanotec, via Amendola 122/D, 70126 Bari, Italy}
\address{$^2$ Department of Chemistry, University of Bari, via Orabona 4, 70126 Bari, Italy}
\address{$^3$ INAF-Osservatorio Astrofisico di Arcetri, Largo E. Fermi 5,
              I-50125 Firenze, Italy}
\address{$^4$ ESA-ESTEC, Aerothermodynamics Section, Kepleerlaan 1, 2200 ag, Noordwijk,
              The Netherlands}

\ead{gaia.miccalongo@nanotec.cnr.it}
\ead{savino.longo@nanotec.cnr.it}
\ead{domenico.giordano@esa.int}

\vspace{10pt}
\begin{indented}
\item[]July 2015
\end{indented}

\begin{abstract}

The diffusion Monte Carlo method with symmetry-based state selection is used to calculate the quantum energy states of H$_2^+$ confined into potential barriers of atomic dimensions (a model for these ions in solids). Special solutions are employed permitting one to obtain satisfactory results with rather simple native code. As a test case, $^2\Pi_u$ and $^2\Pi_g$ states of H$_2^+$ ions under spherical confinement are considered. The results are interpreted using the correlation of H$_2^+$ states to atomic orbitals of H atoms lying on the confining surface and perturbation calculations. The method is straightforwardly applied to cavities of any shape and different hydrogen plasma species (at least one-electron ones, including H) for future studies with real crystal symmetries.

\end{abstract}

\noindent{\it Keywords}: confined H$_2^+$, Monte Carlo methods, plasma-surface interaction, excited states

\ioptwocol

\section{Introduction}\label{intro}

Hydrogen plasma contains ions and radicals which can easily be incorporated into solid lattices. For example, H$_2^+$ is the simplest molecular system and the primary molecular ion produced by radiation or fast electrons in hydrogen gas \cite{mozumder1999,phelps1990,Liebe}. In hydrogen plasma sources, H$_2^+$ is accelerated by sheath fields and hits reactor walls and electrode surfaces with enough energy to remain inside the solid lattice \cite{Dio2,Dio3}. H$_2^+$ in solids have been considered for advanced applications, e.g. H$_2^+$ inside GaAs have been proposed as charge qubits \cite{Shi}. Although in most plasma conditions H$_2^+$ is rapidly converted to H$_3^+$ which becomes the most abundant ion in plasma, this last ion usually dissociates when hitting a surface and it is not found as such in solids \cite{Otte}. H atoms can be produced as high velocity particles e.g. by direct dissociation processes via triplet states or ion-neutral impact and hit solid substrates with residual high speed \cite{Pana, Donko}. Very high energy H or D beams are also produced in the context of fusion reactor experiments \cite{Vanni} providing another case of fast hydrogen species which can be incorporated in structural materials. A similar phenomenology is involved when high energy hydrogen ions hit charged grains in dusty hydrogen plasmas with applications from astrophysics to material science. A better understanding of hydrogen plasma species inside metals and metal hydrides is also relevant for other applications like hydrogen storage \cite{w1,w2}. 

The physical state of hydrogen plasma species, in particular H$_2^+$ and H in solid matter is therefore asking for more investigation.

In this context, the geometrically confined H atom and H$_2^+$ ion are useful first models to describe the state of these species when confined in ionic lattices and metals, as well as models for high pressure plasma conditions \cite{michels1937,degroot1946,yngve1988,yngve1986,aquino2007,micca2,segal2006,soullard2004,
decleva1999,hernandez1998,qiao2002,cottrell1951,singh1964,leykoo1981,goreki1988,mateos2002,sarsa2012,Shi,micca2015} without taking into account details of the confinement field.

The geometrical confinement consists in imposing an artificial nodal surface to the electron wave function, corresponding to the envelope of points where the confining potential reaches infinity value, while the Coulomb potential, due to the nucleus/i, is effective in the region of space enclosed by this nodal surface.

Many previous papers quoted above have considered this problem in the context of quantum chemistry and theoretical physics, but with some limitations which are particularly serious in the perspective of wide application by specialists of other disciplines.

First, typical methods are strongly limited in terms of geometry. In perspective, a method is needed where potential wells of any shape can be considered with little change of the code. This because the wells representing the confinement of real ions can be cubic, square, tetrahedral, spherical while the confining geometry in the literature is usually \emph{ad hoc} instead that related to real applications. For example, for H$_2^+$ the usual choice is a prolate ellipsoidal well, a very artificial geometry: this choice is favored since the wave equation for the electron state factorizes under ellipsoidal coordinates $\xi$ and $\eta$, therefore it is easier to treat a nodal surface of constant $\xi$. 

Second, most available approaches make quite troublesome to consider situations where the protons are not fixed in special positions. Practically this rules out the possibility to calculate IR and Raman spectra.

Finally, it is not simple to consider electronically excited states. These are of course of paramount importance in plasma physics applications, since excited states are connected to optical-UV diagnostics, chemical reactivity of the ion, recombination rates.

Of course detailed calculations for special cases can be found in some of the references given above. What is laking in the interest of the plasma community is a simple approach allowing scientists to develop native code models to be integrated in future multiphysical studies of plasma production of active impurities in solid substrates.

In previous papers by the present authors, it has been found that the application of a simple quantum Monte Carlo method, namely Diffusion Monte Carlo (DMC), allows to include wells of any shape for the case of H atoms \cite{micca2} and allows straightforward calculations of H$_2^+$ $\Sigma$ states with moving nuclei \cite{micca2015}. These results suggest that this approach can be applied generally to excited states of one electron atomic and molecular systems for the above summarized applications. 

This approach, at least for one electron systems like H, He$^+$, H$_2^+$,  HD$^+$, allows effective development of native code.

In this paper we propose a standardized version of our approach as a tool for applications, where hydrogen plasma species in solids are described using geometric confinement.

As an original test case, we provide here new data for the two lowest states of $\Pi$ symmetry, for which actually few information is available even for the free molecule case, under spherical geometry with moving nuclei. This result is obtained by using native {\it fortran} code.

\section{Method} \label{Method}

The approach used is based on the diffusion Monte Carlo (DMC) method \cite{foulkes2001, thijssen1999}.

In this method an ensemble of MC particles or \emph{walkers} performs a diffusion process. This diffusion process occurs in space and in \emph{imaginary time}, which is actually connected to quantum energy. These particles do not represent electrons like in classical MC methods \cite{Longo,Donko2} (although the treatment of the particle list is similar to that reported in these works) but the whole ensemble represents the quantum state of a single electron. An appropriate time step $\tau$ is selected and an iteration process is started: this "time step" is more correctly defined as a squared length.

Regardless of the barrier geometry cartesian coordinates $x,y,z$ can be used, adding simplicity to the code development and flexibility with respect to barrier shape change. Any walker is therefore a pointer $i$ to a list $(x_i,y_i,z_i,f_i)$ where $f$ is a binary flag which can be used to signal if a particle is active or deleted (see below). In the following \textbf{R} is a shortcut for $(x,y,z)$.

At any (imaginary) time step, any walker moves to a new position ${\bf R'}$ which is chosen with probability

\begin{equation}\label{1}
T=\left(2\pi\tau\right)^{-\frac{3}{2}}exp\left[-\frac{\left(\textbf{R}-\textbf{R'}\right)^2}{2\tau}\right]
\end{equation}

The selection in an actual code is done by the rejection technique well known in plasma modeling: three randoms $\eta_1$, $\eta_2$, $\eta_3$ are generated in $(0,1)$ and appropriately scaled e.g. $\Xi_1 =  L(2\eta_1-1)$ where $ L = 4\sqrt{\tau}$. Then $x' = x + \Xi_1$ etc. Another random $\eta_4$ is compared to $p=\exp(-{(\textbf{R}-\textbf{R'})^2)/2\tau})$, if $\eta_4 < p$ the move is accepted, otherwise the process is repeated until \textbf{R'} is accepted.

Note that this formula applies when using atomic units $e = 1$, $m=1$, $h/2\pi = 1$, $4\pi \epsilon_0 =1$. Using these units, the unit of length is Bohr (52.9 pm) and the unit of energy is Hartree (27.21 eV). However the eV has been used in place of this last unit in the plots, following spectroscopy and material science usage.

Any walker can be \emph{deleted} or can \emph{reproduce} itself at the end of the time step. The survival probability is $P$, the reproduction probability is $\max (P-1,0)$ where

\begin{equation}\label{2}
P=exp\left[-\frac{\tau\left(V\left(\textbf{R}\right)-V\left(\textbf{R'}\right)-2E_T\right)}{2}\right]
\end{equation}
where $V$ is the potential energy and $E_T$ is the estimated quantum energy. This last is determined by keeping track of total walker number $N$ using the following expression

\begin{equation}\label{3}
E_{T_i}=E_{T_{i-1}}+\alpha\ln\left(\frac{N_{i-1}}{N_i}\right)
\end{equation}
where $E_{T_{i-1}}$ is the energy value at time step $i-1$ and ${N_{i-1}}$, $\alpha$ is a small energy and $N_i$ are respectively the number of walkers at time step $i-1$ and $i$.

As mentioned, in our simulations, hydrogen plasma species are placed into impenetrable, geometrically symmetric potential barriers.

To implement a barrier is very simple: is a walker enters the forbidden region by diffusion it is deleted. This is consistent with the previous treatment since in the forbidden region $V$ is infinitely high. Otherwise it depends on the system selected.

For example, in the treatment of spherically confined dihydrogen cation H$_2^+$, one eletron and two protons are placed inside an impenetrable sphere radius $r_0$. The two protons (classical) are placed along the x axis, at $x=\frac{d}{2}$ and $x=-\frac{d}{2}$ respectively ($d$ being the internuclear distance). Therefore

\begin{equation}\label{5}
V(d,{\bf r}) = \frac{1}{|{\bf r}-{\bf i}d/2|}+\frac{1}{|{\bf r}+{\bf i}d/2|}+\frac{1}{d}+E_H
\end{equation}

Some comments on this formula. This is not the value in the potential energy curves, this last is $E_T$ which is a function of $d$ and $r_0$. ${\bf i}$ is the x axis versor. It is necessary to include the repulsion energy of the two nuclei which is classical in nature. The addition of $E_H$, which is the ionization energy of the $1s$ state of H, fixes $E_T=0$ as the asymptotic value of a free ($r_0 \rightarrow \infty $) ion at large $d$ (dissociation limit) in the  ground electronic state.

Our approach is simpler than that used in, say, quantum chemistry: we do not use for example the importance sampling technique, which is necessary for many-particle calculations, but adds complexity here. Instead, the initial population of walkers is distributed uniformly in a cube enveloping the barrier. This leads to a sharp start with strong fluctuations in $N$ and $E_T$.

To reduce the fluctuations, an additional solution is implemented in our codes: when the absolute value of the relative variation $(N_{i}-N_{i-1})/(N_{i}+N_{i-1})$ is small enough, in our tests $0.5 \%$, the energy $E_T$ is replaced by the mean calculated in a large previous number of time steps.

This solution is very effective and a stable energy eigenvalues are attained in the limit of long simulation. 

A known drawback of the DMC approach the difficulty to calculate states where the wavefunction changes sign without special solutions \cite{thijssen1999,reynolds1982,umrigar1993}. Even for a single-electron system this problem arises when targeting excited states. The approach suggested here is to target primarily those states whose nodal surfaces are know a priori. The procedure is elaborated in the next section.

To conclude this section, it is appropriate to mention that, specially for relatively simple problems like those arising with hydrogen species, other approaches can be pursued like those based on basis sets (atomic orbitals, gaussians etc.) \cite{aquino2007,ceci}. However, basis sets are not so simply adapted to different shapes of the confining surface. Furthermore, a plasma scientist, as mentioned, will find the a code based on the method in this section highly understandable due to its similarity to a Monte Carlo transport model. This facilitates changes and further development.

\section{Excited states} \label{Excited states}

The DMC method by itself selects the electronic state of minimum energy, e.g. the ground state, for a given nuclei separation $d$.

\begin{figure}
\resizebox{0.5\textwidth}{!}{
  \includegraphics{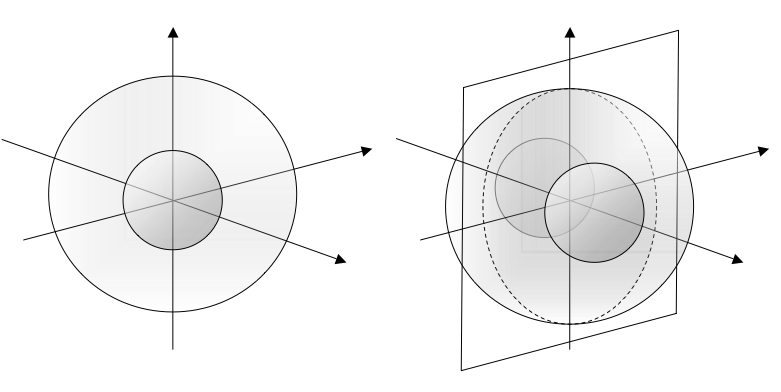}
}
\caption{The use of nodal planes illustrated for a H atom. The $2p$ state (right side) is the lowest energy one under the constraint that walkers cannot cross a plane including the origin. While the $1s$ (left) has no radial node and no nodal plane, $2p$ also has no radial node but it has one nodal plane, the one selected}
\label{nodal} 
\end{figure}

Nodal planes are used in order to select excited states based on their symmetry. For example, the $2p$ atomic state (Figure \ref{nodal}) changes its sign across the plane $x=0$, where z is the coordinate passing through the two nuclei. Therefore, this state can be selected as a target for calculations if walkers are free to propagate only in the half-space $x  > 0$. This result is obtained, based on the previous algorithm, very simply by letting $V$ to assume a very high value when $ x < 0$. Although the wave function is of course negative in the other half-space, it can be shown that quantum mechanical averages can be calculated using only the positive half-space, therefore the sign change is no concern; the only important thing is that the plane $x = 0$ becomes an absorbing barrier to walkers, as described. This plane acts a nodal surface.

Of course this method does not allow to find all excited states, even low energy ones. For example the H(2s) state cannot be found this way. For asymmetric confinements the nodal plane used must be a symmetry plane of the confining surface as well, limiting the choice even further. In case the requested state is not accessible using this straightforward technique alternative solutions within the DMC method are possible (e.g. adaptive search of nodal surfaces) but they are not so simple and basis sets methods may be more appropriate.

In the present calculations, nodal surfaces are used to select the two states $\Pi_u$ and $\Pi_g$. The $\Pi_u$ state is reproduced by placing an extra barrier at the centre of the box, along the $xz$ plane; while, for the $\Pi_g$ state, another barrier has been added along the $yz$ plane.

\section{Test case: $\Pi_u$ and $\Pi_g$ states of spherically confined H$_2^+$} \label{Results}

Little information is still available about the potential energy curve of the low lying excited states of confined H$_2^+$ like $^2\Pi_u$ and $^2\Pi_g$.

This is a very unsatisfactory situation not only in view of the intrinsic importance of these exited states, but also in view of the necessity to access the potential energy curves of low-lying $\Pi$ bound states in future studies of radiation transport and chemical reactivity of confined phases of hydrogen.

In order to get stable and accurate results, we used a time step of $\sim 10^{-3}$ with an average number of walkers $\sim 1000$ and $\alpha = 10^{-3}$ a.u.. A single point requires half an hour on a single node.

\begin{figure}
\resizebox{0.5\textwidth}{!}{\includegraphics{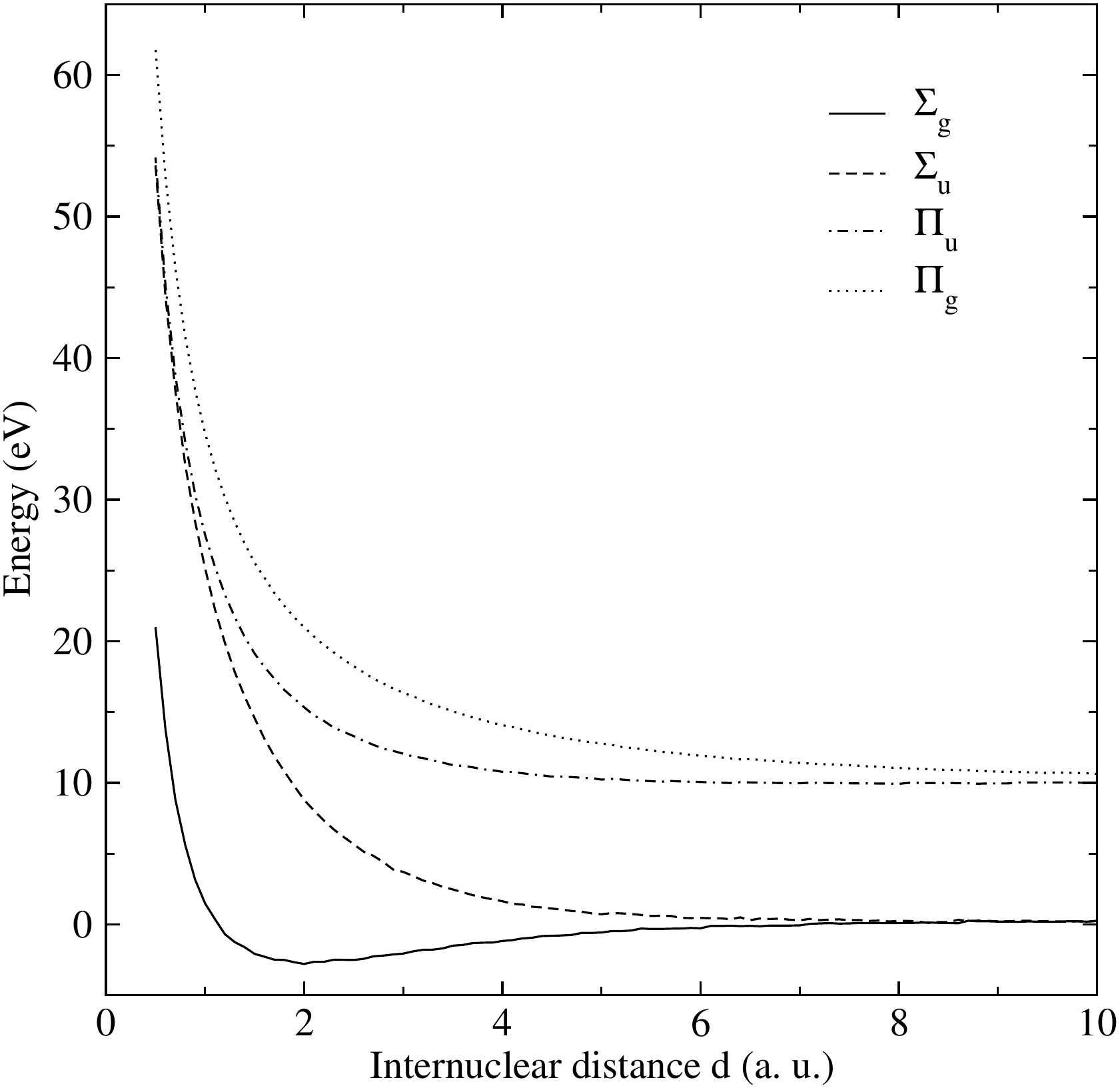}}
\caption{Potential curve of the H$_2^+$ ion $\Sigma$ and $\Pi$ states in diluted gas and plasma phase, calculated using the present code.}
\label{fig:1} 
\end{figure}

In figure \ref{fig:1} the potential energy curves for the $\Sigma$ and $\Pi$ ground states are reported. These two potential energy curves, calculated by our code, are indistinguishable from the best literature results (e.g. \cite{bates} and figure 8.3 in \cite{Liebe}) on the scale of the plot, i.e. they agree with them within 1\% and better results can be obtained for single points at the cost of computer time, by decreasing the time step and increasing the number of walkers. Is should be noted, however, that substantial improvements have a very high computational cost. This is a limit of the DMC approach and other methods should be considered when high precision is needed.

\begin{figure}
\resizebox{0.5\textwidth}{!}{\includegraphics{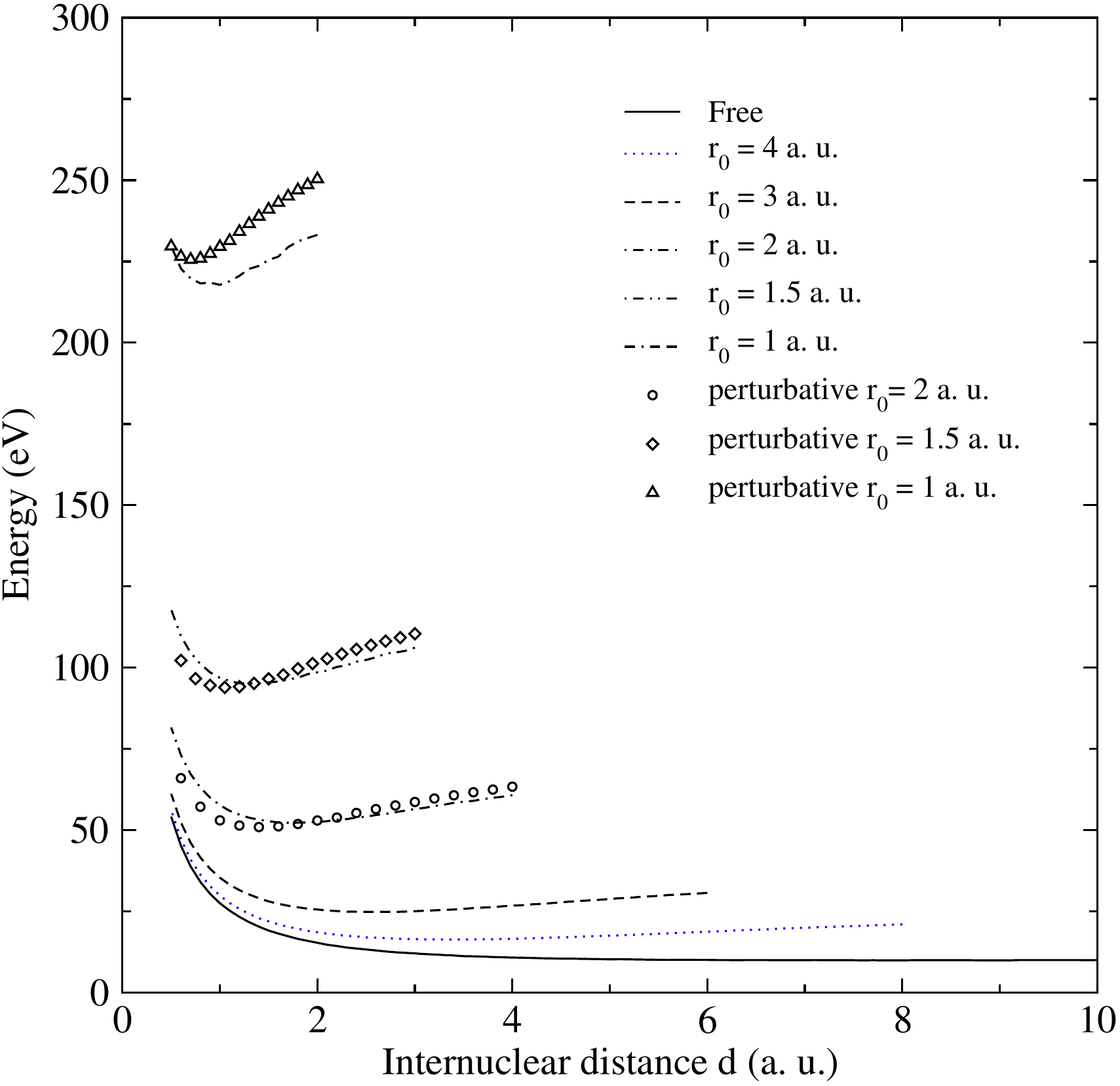}}
\caption{Potential curve of the $\Pi_u$ state of a H$_2^+$ molecule confined into a spherical well. A few perturbation calculations are included in the same plot.}
\label{fig:2} 
\end{figure}

\begin{figure}
\resizebox{0.5\textwidth}{!}{\includegraphics{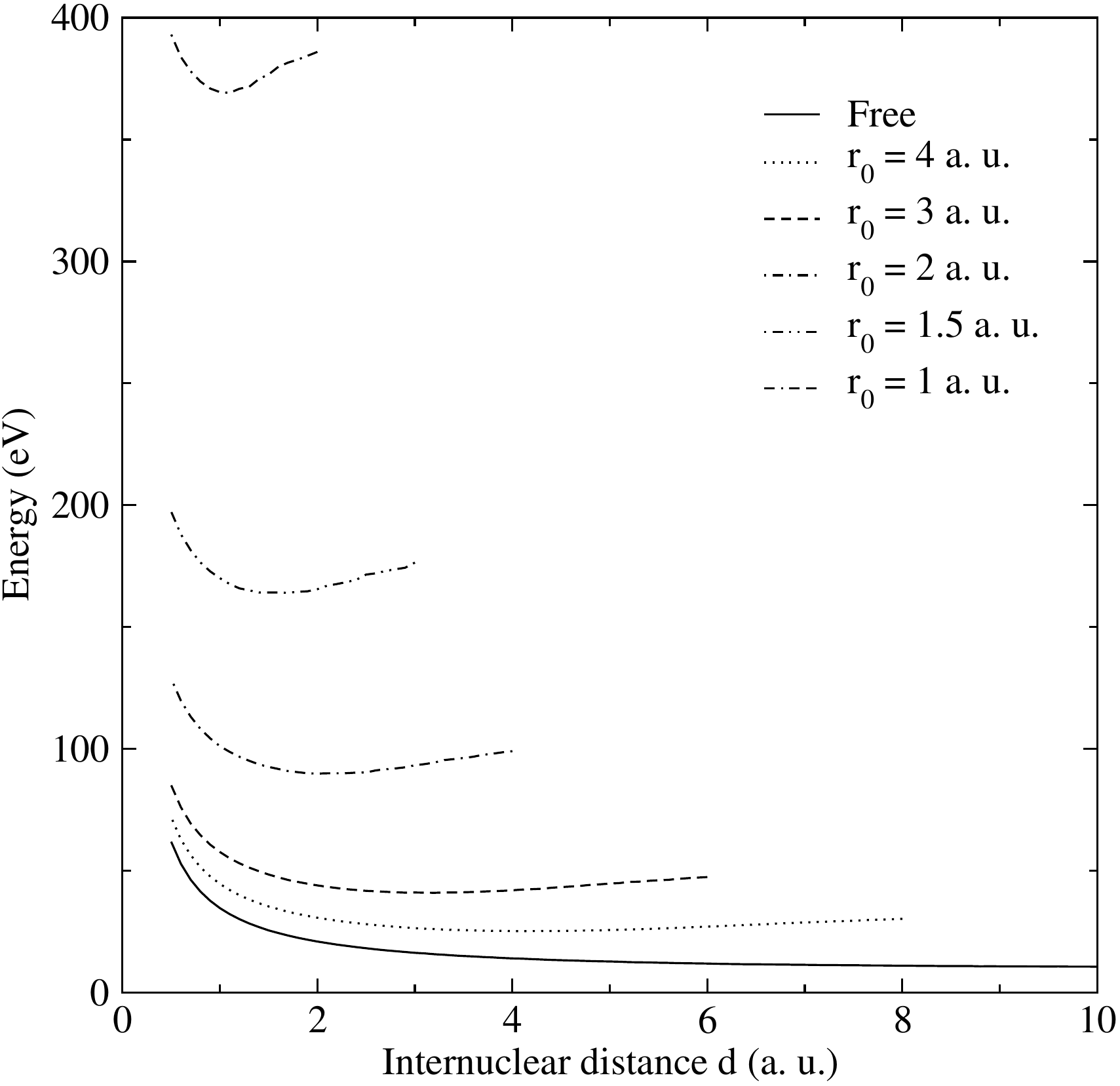}}
\caption{Potential curve of the $\Pi_g$ state of a H$_2^+$ molecule confined into a spherical well}
\label{fig:3} 
\end{figure}

Figures \ref{fig:2} and \ref{fig:3} show the potential energy curve calculated for for $\Pi_u$ and $\Pi_g$ states respectively, for different values of the confining sphere radius.
It was shown in \cite{micca2015} that confinement changes the character of the state of $\Sigma_u$ symmetry, which develops a pronounced minimum which is absent in the free molecular state. 

\begin{figure}
\resizebox{0.5\textwidth}{!}{\includegraphics{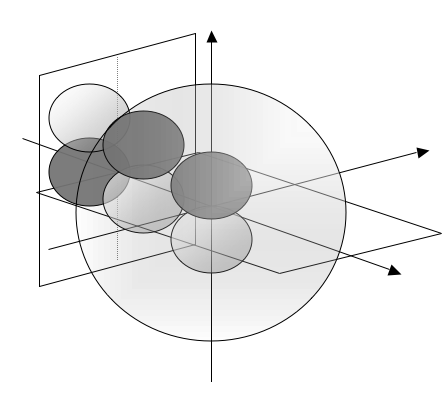}}
\caption{Sketch of the correlation of $\Pi_u$ to 3d at the sphere surface. Note the two nodal planes: one selecting the initial state and the second one mimiking the effect of the sphere surface for large $r_0$.}
\label{fig:4} 
\end{figure}

Based on the reported results, a comparison of the reaction of $\Pi$ and $\Sigma$ states of H$_2^+$ to compression is possible. The bound $\Pi$ state, as known, presents a very shallow potential minimum. In this case, therefore, the change in character due to compression is much more important. Furthermore, even mild compression which barely affect $\Sigma$ states has an effect here, due to much larger internuclear distance.

The qualitative features of the new states can be further discussed using orbital correlation \cite{harris} a standard tool in chemical physics: we will show that a version of this technique, which allows a quantitative discussion of H$_2^+$ energy for limiting cases, is suggested by the use of planes as nodal surfaces in the model itself.

\begin{figure}
\resizebox{0.5\textwidth}{!}{\includegraphics{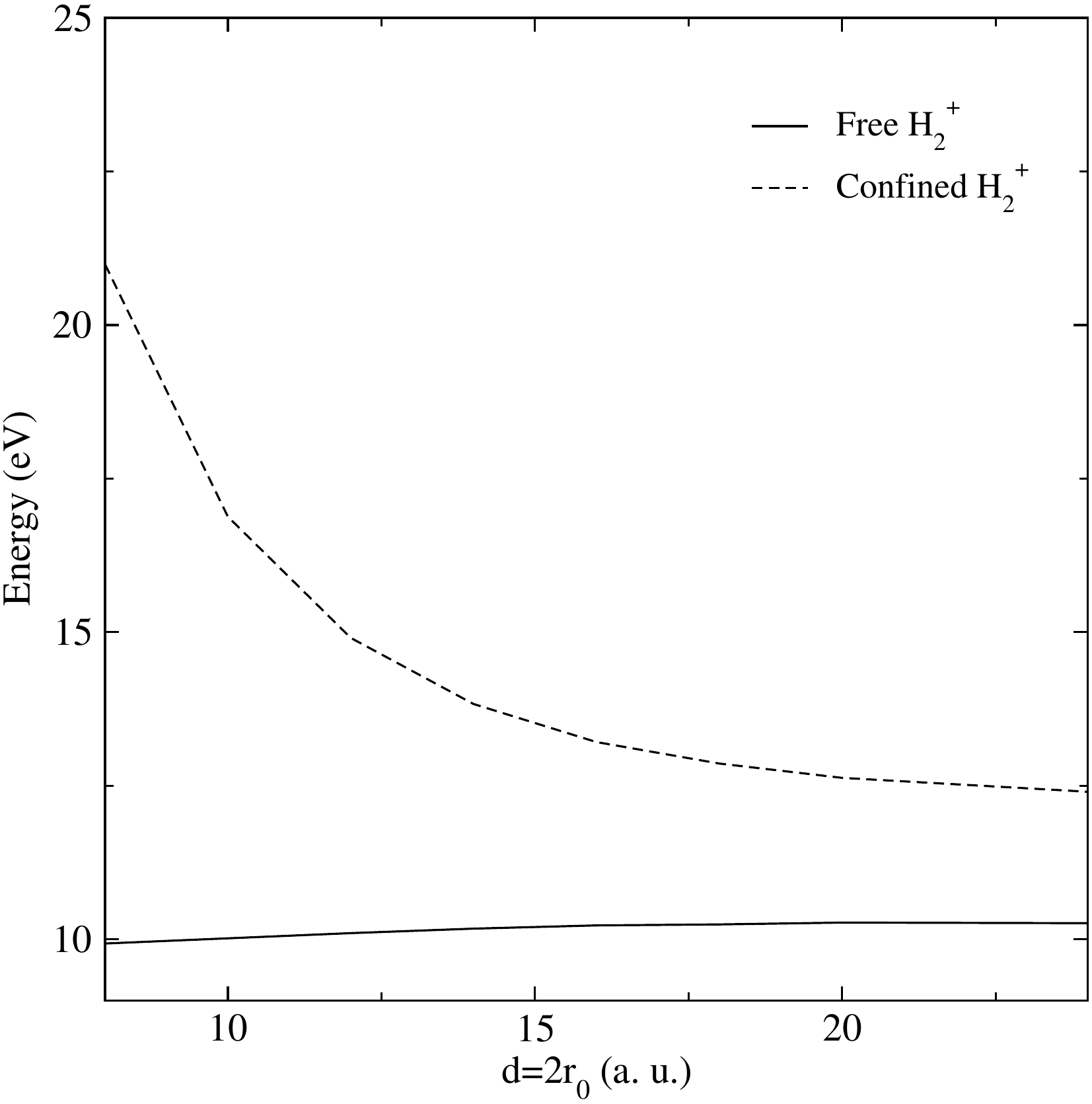}}
\caption{Total energy of the free and confined $\Pi_u$ states of H$_2^+$ as a function of d for variable $r_0 = d/2$. Upper curve illustrates the limit case of highest nuclear separation.}
\label{fig:5} 
\end{figure}

At the lightest degrees of compression, the increase of the energy value for $d\rightarrow2r_0$ is explained by the compression of the separate atomic states. 
In the case of mild compression, it is sound to consider the separate atom states are $2p$ states atomic states for both for $\Pi_u$ and $\Pi_g$. When the nuclei reach, at the same time, the sphere surface, this surface acts as a further node leading to $3d$ atomic states of higher energy (figure \ref{fig:4}), this trend is also evidenced by the total energy trend as a function of d for variable $r_0 = d/2$. The result for $^2\Pi_u$ is shown on figure \ref{fig:5} and it can be noticed that
the correlation mechanism accounts for the quantitative features of the potential energy surface for large
values of $r_0$. Note in fact that, while the lower curve tends to a limit given by the $2p-1s$ H orbital energy (10.2 eV), the upper one tends to $3d-1s$ (12.1 eV) due to the confinement.

The compression by a smaller 
sphere produces, like in the case of $\Sigma$ states \cite{micca2015}, a shorter 
equilibrium distance of the two nuclei. This effect is due to the increase of electron density in the sphere volume. This can be checked 
by an even simpler method for very approximate estimates, a first-order perturbation calculation given by

\begin{equation}\label{6}
E(d,r_{0}) \sim E_{110} - <\frac{2}{|\textbf{r}-{\textbf{i}}\frac{d}{2}|}>_{110} +
\frac{1}{d} + E_H
\end{equation}
 
where the average is based on the analytical wave function of the excited ${110}$ state for a particle in a spherical well. Results of this formula are also reported in figure \ref{fig:2}, providing some validation (or course the DMC results are better, specially for mild confinement) and showing that the minimum in the potential energy curve is explained as the effect of coulomb attraction between protons and an essentially free electron cloud.
 
The change in character of the $\Pi$ states suggests that radiation diffusion mechanism by H$_2^+$ under strong pressure and involving these states is qualitatively affected, with potential impact on UV absorption and emission of H$_2^+$ confined in solids.

Although not demonstrated here (but already shown by the same present authors for the case of confined H \cite{micca2}), these codes are readily modified to consider e.g. cubical cavities, therefore this approach opens a wide range of research possibilities.

\section{Conclusions}\label{conc}

In this paper we report recent results obtained by the authors to develop and use a simple, versatile method for the determination of quantum states of hydrogen plasma species confined inside potential wells.

Our method consists in applying a basic version of the DMC method with a special solution to adapt the energy: this approach allows to get results with a minimum of development effort and very high adaptability.

As a test case, we demonstrate here the approach for two important low-energy excited states, namely $\Pi_u$ and $\Pi_g$ for spherical confinement: approximate calculations and correlation considerations are used to confirm the validity of these result.

Results like those produced by the approach proposed can be very valuable in a multi-physical models of hydrogen plasma reactors. For example, the vibrational levels of confined H$_2^+$ in different electronic states can be obtained based on the potential energy surface provided - based on these vibrational levels calculations of chemical reactivity, visible and Raman spectra become accessible. These may be compared to future measurements to infer the physical state of ions inside reactor materials and substrates.

In this paper we have applied the method to "hard" confinement (e.g. infinite potential barrier) which is the most simple version of confinement and probably the best start, with a minimum of parameters. As shown in literature, several variations of the concept allow to describe different systems. For example, a "soft" confinement \cite{colin2011} is based on a finite barrier and describes confined systems which may undergo diffusion between confining sites, while smooth potential wells \cite{connerade1999, connerade2001} are more suitable to describe specific cages or dense plasmas (confinement in the Debye sphere). The method described in this paper is easily adapted to these variations. Its efficacy in such cases, however, needs to be investigated.

In perspective, the method used is appropriate to many confinement cases in view of the increasing range of applications, allowing to produce in the long term a variety of results of the quantum behavior of species of interest of the plasma and material processing communities.


\section*{Acknowledgment}
This research activity has been supported by the General Studies Programme of 
the European Space Agency through contract 4200021790CCN2.

\end{document}